\documentclass[letterpaper,aps,floatfix,twocolumn,showpacs]{revtex4}
\usepackage{graphicx}
\usepackage{epsfig}     
\usepackage{amsmath}
\usepackage{commath}
\usepackage{hyperref}

\usepackage{setspace}
\newcommand{\old}[1]{{}}

\begin{document}

\title{A crisis for the V\&V of turbulence simulations} 

\author{James Glimm}
\email{glimm@ams.sunysb.edu}
\affiliation{Stony Brook University, Stony Brook NY 11794}
\author{Baolian Cheng}
\email{bcheng@lanl.gov}
\affiliation{Los Alamos National Laboratory, Los Alamos NM 87545}
\author{David H. Sharp}
\email{dcso@lanl.gov}
\affiliation{Los Alamos National Laboratory, Los Alamos NM 87545}
\author{Tulin Kaman}
\email{tkaman@uark.edu}
\affiliation{University of Arkansas, Fayetteville AR 72701}


\begin{abstract}
Three very different algorithms have been proposed for solution of the
Rayleigh-Taylor
turbulent mixing problem. They are based upon three different 
physical principles governing the Euler equations for fluid flow,
which serve to complete these underspecified equations
by selection of the physically relevant solution from among the many
otherwise nonunique solutions of these equations.
The disputed physical principle is the admissibility condition which selects the
physically meaningful solution from among the myriad of nonphysical solutions.
The three different
algorithms, expressing the three physical admissibility  principles,
are formulated alternately
in terms of the energy dissipation rate
or the entropy production rate.
The three alternatives are zero, minimal or maximal
rates. 
The solutions are markedly different. We find strong validation evidence
that supports the solution with the maximum rate of dissipated energy,
based on a review of prior results and new results presented
here. Our verification reasoning, consisting of
mathematical analysis based on  physics assumptions, also supports the
maximum energy dissipation rate and reasons against the other two. 
The zero dissipation solution is based on claims of direct numerical 
simulation. We dispute these claims and introduce analysis indicating
that such simulations are far from direct numerical simulations.

Recommendations for the numerical modeling of the deflagration to detonation
transition in type Ia supernova are discussed.
\end{abstract}

\maketitle


\section{The crisis for V\&V}
\label{sec:crisis}

The solutions of the Euler equation for fluid dynamics are not unique.
An additional physical principle in the form of an admissibility criterion is
needed to select the physically meaningful solution. Wild and manifestly
nonphysical solutions have been studied extensively \cite{DelSze09,DelSze10}.
These constructions offer cautionary 
counter examples to enlighten studies of the Euler equation as a model for
fully developed turbulence. This paper is concerned with
less dramatic, and in that
sense more troublesome examples of the nonuniqueness
for Euler equation solutions: ones
that are the limit of mesh generated solutions as
the mesh tends to zero. 

We identify three different algorithms with markedly different
solutions, based on three different physical principles of admissibility.
The primary distinction 
among the three is in 
the grid level dissipation of energy: zero, minimal or maximal.
We find strong evidence in support of the maximal dissipation of
energy, which we formulate as the admissibility condition. This
admissibility condition is the physical principle needed to
complete the Euler equations as a model of fluid turbulence.
The main thrust of this paper is the 
introduction of new evidence and the review of existing evidence to
justify the maximum rate of energy dissipation admissibility condition.
We believe that, equivalently, the maximum rate of entropy production
is a correct admissibility to complete the definition of the Euler equations
as a model of fully developed turbulence.

The nature of the problem is summarized in the quote from the comprehensive
survey articles \cite{Zho17a,Zho17b}
regarding turbulent mixing for acceleration driven flows, from
\cite{Zho17a}, Sec.~ 6
regarding evaluation of the Rayleigh-Taylor (RT)
instability growth rate $\alpha_b$,
``agreement between simulations and experiment are worse today
than it was several decades ago because of the availability of more
powerful computers.''
The quote defines a challenge to existing standards of verification and
validation (V\&V). 
While the solution is disputed at the level of physics, while the
governing physical model is in dispute, the crisis persists.
The evidence presented here, supporting the maximum rate of dissipation
admissibility, falls within the general framework of V\&V.

The experimental
evidence \cite{SmeYou87} in agreement with the  physical principle of
maximum rate admissibility condition is a validation of the 
maximum rate algorithm that is derived from this
law.  The mathematical proof introduced in Sec.~\ref{sec:p=1}
to deduce the missing physical principle, a verification step,  is 
derived using physically motivated assumptions. This 
reasoning shows that the maximum rate admissibility condition is
an intrinsic aspect of physically meaningful simulations of turbulence.

Pending a resolution of the admissibility issues,  the crisis for
the V\&V of numerical simulations of turbulence remains. It seems that the
crisis does not pertain to the V\&V methodology, but rather to the
lack of its application to this problem.

We are not aware of other verification tests to
distinguish among the alternate  admissibility laws of physics.
To our knowledge, the body of
validation tests for turbulence simulations does not 
distinguish among the alternate physical principles of admissibility
conditions. To illustrate this point, the
hot-cold water channel experiments of \cite{MueSch1_09} match simulation
data for all three algorithms, and thus does not distinguish among
the three physical principles. The more demanding salt-fresh
water channel, which likely would have differentiated among the
algorithms, was not considered in a validation study of the zero or
minimum dissipation algorithms. The analogous salt-fresh water experiment
Exp. 112 of \cite{SmeYou87}, with exceedingly tight experimental
error bars, provided validation data for simulations \cite{LimKamYu12} of
the maximum energy dissipation algorithm and physical principle.
Further comments on this matter are found in Sec.~\ref{sec:scaling}.

We believe that this paper contributes a 
resolution of the V\&V crisis, in favor of the maximum rate laws for
admissibility of fluid turbulence.

If turbulence or nonturbulent stirring
is present in the problem solved, we propose that
{\it{the standards of V\&V
should ensure the physical relevance of the solutions obtained.}}

\subsection{The three disputed physical principles}
\label{sec:3laws}

The three physical principles lead to three distinct algorithms,
which differ in their treatment of mesh level dissipation
of energy, from zero to limited to full (maximum) dissipation rates.
The more recent zero dissipation algorithm shows the largest
disparity relative to experiment. The limited dissipation algorithm
is also in disagreement, but less so, thus
providing the basis for the comment of Zhou.

The evidence to choose among these three physical principles
is based on: (a) numerical simulations in agreement with experiments
(validation) and (b) mathematical proof (with physically
motivated hypotheses) of the admissibility criterion
(verification).

The extensive prior experimental validation
evidence with data from \cite{SmeYou87}
is reviewed in summary form in Sec.~\ref{sec:scaling}.
A new simulation showing comparison of scaling law exponents is also
presented in Sec.~\ref{sec:scaling}. 
Both strongly favor the maximum
dissipation rate.
For now we quote from Zhou \cite{Zho17a}, Sec. 5.2, in discussing 
the solution with the maximum rate of energy dissipation \cite{GeoGliLi05}:

``it was clear that accurate numerical tracking to control numerical mass 
diffusion and accurate modeling of physical scale-breaking phenomena and
surface tension were the critical steps for the simulations to agree with
the experiments of Read and Smeeton and Youngs''.

To account for observed discrepancies between predictions 
based on the minimum dissipation algorithm (Implicit Large Eddy Simulation
 or ILES) and 
experimental data, it is common to add ``noise'' to the physics model.
As noise increases the entropy, some discrepancies between simulation
and measured data are removed. 
We have seen \cite{GliShaKam11,ZhaKamShe18} that the noise
in the experimental data \cite{SmeYou87} is 
insufficient to restore agreement with data. It
accounts for at most a 5\% effect relative to  the
experimental data \cite{SmeYou87} used for validation.
Both the zero and the minimum
dissipation physical principles and algorithms are 
in disagreement with the experimental data of \cite{SmeYou87}.

The mathematical evidence, developed in 
Sec.~\ref{sec:p=1},
consists of a mathematical
proof of the maximum energy dissipation rate principle for turbulent
flow based on physically motivated hypotheses. This is a verification step.
This evidence
strongly favors the maximum dissipation rate physical principle.

A mathematical proof of the principle of maximum entropy production
can be derived from laws of statistical physics applied to molecular
physics \cite{Lan73,Leb78},
based on the thermal fluctuations of particle positions.
This analysis leads to the phenomenological Fourier law of 
thermal conductivity. 
The use of a maximum entropy production admissibility principle is familiar from
numerical modeling of shock waves. This entropy is the thermal
entropy of the molecules. 

The entropy related to turbulence and its intermittency
concerns random fluctuations of the particle velocities, rather than
their positions.
A mathematical proof of the principle of
maximum entropy production rate, extended to particle velocities rather
than positions, (verification) 
requires revisiting and extending this classical analysis. 
Thus mathematical support for the maximum entropy rate
verification principle is suggestive, not definitive.

The maximum rate of entropy production, as an admissibility condition 
for fully developed turbulence, is an extension of the second law of 
thermodynamics, in the sense that under this extension,
the physically admissible dynamic processes are constrained
more tightly than those allowed by the second law itself.
This principle  has
 been applied successfully to many natural processes 
 \cite{MarSel06,MihFarPai17} 
including problems in climate science (terrestrial and other planets) \cite{OzaOhmLor03}, in
astrophysics, and the clustering of galaxies. As
noted in \cite{KleDyk10}, it does not have the status of an
accepted law of physics. Arguments in favor on maximum
principles (entropy production or energy dissipation) can be found in
\cite{GeQia09,MihFarPai17}.

\subsection{The three algorithms}
\label{sec:3-algo}

The maximum dissipation algorithm is
FronTier. It is
based on dynamic subgrid scale models (SGS) in the spirit of
\cite{GerPioMoi91,MoiSquCab91},
and front tracking. It is the algorithm referenced in the
second quote from Zhou above.

Reynolds averaged Navier Stokes (RANS) simulations resolve all length
scales needed to specify the problem geometry.
Large eddy simulations (LES) not only
resolve these scales, but in addition they resolve some, but not 
all, of the generic turbulent flow. The mesh scale, i.e., the finest of the
resolved scales, 
occurs within the turbulent flow. As this is a strongly coupled flow
regime, problems occur at the mesh cutoff. Resolution of all relevant
length scales, known as Direct Numerical Simulation (DNS) is 
computationally infeasible for many problems of scientific and
technological interest. 

The limited dissipation algorithm is ILES. We refer to the algorithm
with no dissipation modeled at all as macro defined DNS (MDNS),
in that the DNS criteria in this algorithm 
are specified in terms of macro flow quantities.  MDNS falls well 
short of true DNS, according to estimates of Sec.~\ref{sec:mdns}.

\subsection{Subgrid scale terms}
\label{sec:sgs}

The subgrid scale
(SGS) flow exerts an influence on the flow at the resolved level.
Because this SGS effect
is not part of the Navier-Stokes equations,
additional modeling terms are needed in the LES discretized equations. These
SGS terms added to the right hand side (RHS) of the
momentum and species concentration equations
generally have the form
\begin{equation}
\label{eq:sgs}
\nabla\nu_t \nabla \quad  {\mathrm{and}} \quad \nabla D_t \nabla \ .
\end{equation}
The coefficients $\nu_t$ and $D_t$ are called eddy viscosity and eddy
diffusivity.

According to ideas of Kolmogorov \cite{Kol41}, the energy in a turbulent
flow, conserved, is passed in a cascade from larger vortices to smaller ones.
This idea leads to the scaling law \cite{Kol41}
\begin{equation}
\label{eq:K41}
\langle |v(k)|^2 \rangle = C_K \epsilon^{2/3} |k|^{-5/3}
\end{equation}
for the Fourier coefficient $v(k)$ of the velocity $v$. Here
$C_K$ is a numerical coefficient and
$\epsilon$, the energy dissipation rate,  denotes the rate at which the energy
is transferred within the cascade from the large scales to the smaller ones.
It is a measure of the intensity of the turbulence.

At the grid level, the numerically modeled 
cascade is broken. Energy accumulates at the grid level. The role of the
SGS terms is to dissipate this excess grid level energy so that the
resolved scales see a diminished effect from the grid cutoff.
This analysis motivates the SGS coefficient $\nu_t$, while a conservation
law for species concentration
similarly motivates the coefficient $D_t$.

ILES is the computational model
in which the minimum value of $\nu_t$ is chosen so that a minimum of  grid level
excess energy is removed to retain the $|k|^{-5/3}$ scaling law.
The prefactor $C_K\epsilon^{2/3}$ is not guaranteed. ILES
depends on limited and globally defined SGS terms. It does not use 
the subgrid terms that correspond to the
local values of the energy dissipation cascade. Miranda is a modern
compact scheme. 
An ILES version of Miranda is presented in \cite{MorOlsWhi17}. This reference
provides details for the ILES construction and an analysis of
scaling related properties of the RT solutions the 
algorithm generates. The subgrid terms are chosen
not proportional to the Laplacian as in (\ref{eq:sgs}),
but as higher power dissipation rates,
so that large wave numbers are more strongly suppressed. 
The SGS modeling coefficients $\nu_t$ and $D_t$ are chosen as global
constants. The basis for the choice is to regard the accumulation of
energy at the grid level as a Gibbs phenomena to be minimized 
\cite{MorOlsWhi17}. Miranda
achieves the ILES goal of an exact $-5/3$ spectral decay,
see Fig. 3 right frame in Ref. \cite{MorOlsWhi17}.

FronTier uses dynamic 
SGS models \cite{GerPioMoi91,MoiSquCab91},
and additionally uses a sharp interface model to reduce numerical
diffusion. In this method, 
SGS coefficients $\nu_t$ and $D_t$ are defined in terms
based on local flow conditions,
using turbulent scaling laws, extrapolated from an analysis
of the flow at one scale coarser, where the subgrid flow is known.
The MDNS algorithm omits all SGS terms completely.

The philosophy and choices of the SGS terms are completely different among
MDNS, ILES and
FronTier. This fact originates in differences in the disputed physical 
principle of admissibility,
and leads to differences in the solutions obtained.
Solution differences between FronTier and ILES were reviewed in
\cite{ZhaKamShe18}, with FronTier but not ILES showing agreement with the
data \cite{SmeYou87}. The MDNS schemes totally lacking SGS terms are even 
further from experimental validation. No physical principle has been advanced
to motivate the MDNS choice of zero dissipation. It appears rather
to result from a belief that MDNS is true DNS and as such,
needs no subgrid terms.

There is some support for this belief, in that a number of authors
label as DNS, simulations in which the $\Delta x$ equals or is even 
modestly larger than the macro defined Kolmogorov scale. It seems to be
a universal practice, however,
in such studies, to include comparison to experiment
relative to the quantity of interest. In other words, the use of
MDNS is accompanied by a validation study. For \cite{CabCoo06}, 
this standard is not followed, and the results are invalidated by experiment.
The refs. \cite{KanMor07,SawYeu15} focus on the
local turbulent intermittency at the MDNS defined scale, and finds a range of
power law behavior, indicating that the MDNS scale is in fact not DNS.

\subsection{MDNS vs. DNS}
\label{sec:mdns}

True DNS of fluid mixing flows means full resolution
of all flow variables. This goal, which means $\Delta x \ll \eta$,
with $\eta$ the Kolmogorov scale,
for DNS relative to the viscosity, and further refinement 
to the Batchelor scale if the
problem Schmidt number $Sc < 1$, is prohibitive in computational cost,
and is not achievable for most 
meaningful problems. The goal of DNS is to avoid the
ambiguities of the SGS terms, and compute in a reliable, and model-free
manner, with no use of SGS terms.
To achieve this goal, a number of compromises  have accumulated
under the same terminology of DNS. In the analysis of turbulent boundary layers,
where the proper definition of the SGS terms is still open, values as large as
$\Delta x \sim 15 \eta$ are considered \cite{MoiMah98}, but in recognition
of the liberties taken with the definition of DNS, validation against 
experimental conclusions is included. DNS is also popular within the
turbulent mixing community, and again there are compromises involved in the
usage of this term.

Proponents of this use of DNS point to possible difficulties in the
experimental data, and do not follow \cite{MoiMah98} with
an experimental validation of their conclusions. We did not locate a precise
definition of DNS for  \cite{CabCoo06} and so we appeal to a similar terminology
regarding a similar code, Miranda \cite{RehGreOls17}. In essence, the
compromise is a substitution of globally defined variables for 
the true DNS choice of local ones, as we now explain. 

MDNS is analyzed in terms of global flow quantities, so that
\begin{equation}
\label{eq:global}
Re_{\mathrm{global}} = \delta v_{\mathrm{global}} L/ \nu \ ,
\end{equation}
where $\nu$ is the kinematic viscosity, $L$ 
is the domain size, with the value $L = 2 \pi l = 2 \pi \times 3072$
in the notation of \cite{CabCoo06}.
$\delta v$ is a velocity fluctuation,
$l$ is the number of MDNS mesh zones in a 
single dimension and the mesh size $\Delta x$ has the value
$\Delta x_{\mathrm{MDNS}} = 1$ in the notation of \cite{CabCoo06}.
From these quantities, we compute, with parameter values 
from \cite{CabCoo06},
\begin{equation}
\label{eq:MDNS-local}
Re_{\mathrm{MDNS}} =
\delta v_{\mathrm{MDNS}} \Delta x_{\mathrm{MDNS}} /\nu
= \eta_{\mathrm{MDNS}} = 1 \ .
\end{equation}

An MDNS specification of these quantities depends on 
$\delta v_{\mathrm{MDNS}}$. In the Miranda definition of LES, 
global SGS terms (eddy viscosity and eddy diffusivity) are chosen.
The dynamic SGS models \cite{GerPioMoi91,MoiSquCab91} are dependent on
local flow properties and are used in the FT algorithm. These
dynamic SGS models contrast to 
the global Miranda definitions of LES, and also to
the SGS eddy viscosity and DNS,
(see eq.~(32) of \cite{RehGreOls17}). In this equation, it is required
that the globally defined eddy viscosity should be small relative
to the physical viscosity.
This globally based
definition implicitly fixes the quantity $\delta v_{\mathrm{MDNS}}$
in terms of globally defined, not locally defined fluctuations.

In \cite{CabCoo06}, pg. 563, it is stated that $\eta = \Delta$, with $\Delta$
the mesh size. The
assumption that $\eta$ is determined from globally defined variables is
clear from the context. Similarly, \cite{CabCoo06} reports
a grid level $Re$ of about 1, again
using macro, not local flow parameters to define $Re$.
Thus we conclude that \cite{CabCoo06} presents an MDNS algorithm.

In terms of the DNS quantities, the mesh scale Reynolds number and Kolmogorov
scale are defined as
\begin{equation}
\label{eq:DNS-RE}
Re_{\mathrm{DNS}} = \delta v_{\mathrm{DNS}} \Delta x_{\mathrm{DNS}} / \nu
= \eta_{\mathrm{DNS}} \ll 1 \ .
\end{equation}
DNS specifies $\delta v$ so that (\ref{eq:DNS-RE}) is satisfied, 
with $\Delta x_{\mathrm{DNS}}$ unknown. This unknown quantity $\Delta x$
prevents detailed analysis of
the incremental  mesh refinement needed for DNS, beyond MDNS.
The inequality (\ref{eq:DNS-RE})
is required at every mesh cell, so that peak local fluctuations (intermittency
related) in the velocity satisfy
this definition. 
\old{
In contrast, the use of macro 
defined flow variables, such as a macro defined Reynolds number,
Kolmogorov scale $\eta$, and SGS  terms, when used in formulas such as 
(\ref{eq:global},\ref{eq:MDNS-local}) give rise to what we have called MDNS. 
These formulas miss the mesh level variations in $\delta v_{\mathrm{MDNS}}$. 
}

The relevant norm to assess the velocity fluctuations is $L_\infty$. This
norm is inconvenient to work with theoretically. Thus we underestimate the
mesh level DNS requirements by considering velocity fluctuations in $L_1$,
where we compute the scaling properties in terms of a length scale parameter 
$l$.  Even with the $L_1$ estimate, intermittency
related local turbulence forces extensive refinement beyond the
nominal (MDNS) level, as is well known.

We estimate the $l$ scaling of the added computational cost
by appeal to structure functions, which we now define.
The structure functions  make precise the intuitive picture 
of multiple orders of clustering for intermittency.
There are two families of structure functions, one for 
velocity fluctuations  and the other for  the energy
dissipation rate $\epsilon$. The structure functions are the expectation
value of the $p^{th}$ power of the variable.  Each has an averaging radius
$l$, and this gives rise to asymptotic scaling as a power of $l$.
For each value of $p$,
the structure functions define a fractal related to their
power law in their decay in the scaling 
variable $l$. The structure functions and the associated scaling exponents
$\zeta_p$ and $\tau_p$ are defined as
\begin{equation}
\label{eq:zeta-tau}
\langle \delta v_l^p \rangle \sim l^{\zeta_p}
\quad {\mathrm{and}} \quad
\langle \epsilon_l^p \rangle  \sim l^{\tau_p}
\end{equation}
where $\delta v_l$ and $\epsilon_l$ are respectively
the averages of velocity differences and
of $\epsilon$. The averages are taken differently to allow for the
systematic differences in signs, $\zeta_p \ge 0$ and $\tau_p \le 0$.
The $\epsilon$ average is over a ball $B_r(x)$ of radius $r$ centered at $x$.
We consider the individual tensor contributions $\delta_j v_i$ 
to (\ref{eq:zeta-tau}). In defining $\zeta_p$ and $\epsilon_p$, we
sum over all tensor indices
$i,j$. The tensor velocity difference is defined as 
$|(\delta_{+,i} + \delta_{-,i}) v_j| / 2$, where $\delta_{\pm,i}$ is the
forward (backward) difference in the coordinate direction $i$
with a step size $l$.
As a special property of the power $p = 1$, we note that 
a change in order of integration eliminates the average over $B_l$, so that the
global average of $\epsilon(x,t)$, which we denote
$\epsilon_0(t) = \int_V \epsilon(x,t) \dif x$ is finite for a.e $t$.
The two families of exponents are related by a simple scaling law
\begin{equation}
\label{eq:zeta-tau2}
\zeta_p = p/3 + \tau_{p/3}
\end{equation}
derived on the basis of scaling laws and dimensional analysis \cite{Kol62}.

We are not aware of
quantitative predictions for the mesh requirements of (true)
DNS simulations, which depend on an estimate of $\delta v_{\mathrm{MDNS}}$.

To summarize this discussion, we state that MDNS can be regarded as an 
exceedingly fine LES. As such, it has a need for its own SGS terms, often
ignored, and for its own validation tests, normally followed, but
in the case of \cite{CabCoo06} actually invalidated. 
The actual level of incremental mesh refinement
needed to achieve true DNS from an
MDNS mesh is a research question.
A definitive test for true DNS is to insert
the dynamic SGS terms into the putative true DNS simulation,
and observe that their value is negligible in the $L_\infty$ norm.
This or any other test of a true DNS simulation is missing in \cite{CabCoo06},
and it is fair to state that the DNS claims of \cite{CabCoo06}
are unsubstantiated.

Returning to the
quote from Zhou \cite{Zho17a}, it seems that the degradation in
validation for RT experiments results not from the more powerful
computers now in use, but from an invalid scientific methodology in their use.

\section{Instability growth rates and scaling laws
compared for three algorithms}
\label{sec:scaling}

We refer to the Rayleigh-Taylor (RT) unstable turbulent mixing
process and characterize in summary the principal differences in
the instability growth rates obtained from the three
proposed laws of physics and the resulting three algorithms. 

RT unstable flow is generated experimentally \cite{SmeYou87} by taking a 
tank, with light fluid above the heavy (stable to gravity), and accelerating it
rapidly downwards, thereby reversing the gravitational and inertial forces.
The resulting flow is unstable and a mixing layer grows on an acceleration
($t^2$) time scale, according to the formula describing the
penetration $h_i$ of the each of 
two fluids into the dominant phase of the other,
\begin{equation}
\label{eq:alphab}
h_i = \alpha_iAgt^2 \ .
\end{equation}
Here $i = 1,2$ denotes the heavy or light fluid, $g$ is the reversed 
acceleration force, and the Atwood number 
$A = (\rho_1 - \rho_2)/(\rho_1 + \rho_2)$
is a buoyancy correction to $g$. $\rho_i$ denotes a fluid density.
It is common to refer to $2 = b$ penetrations as bubbles.

We summarize in Table~\ref{table:compare}
the major code comparisons of this paper, 
based on the RT instability growth rate $\alpha_b$. More detailed comparisons
are found in \cite{ZhaKamShe18,GliShaKam11,LimIweGli10} and references cited
in these papers.
An MDNS scheme, compact and  higher order \cite{CabCoo06} has the
smallest value $\alpha_b$. ILES is larger, and the FronTier
scheme using dynamic SGS is the largest of the three. Among the three 
algorithms, FronTier is
uniquely in agreement with experiment. We have already noted the absence of
noise in the data \cite{SmeYou87}. For problems which are
(a) intrinsically noisy, (b) diffusive, (c) weakly turbulent, (d) 
limited in the objective functions used for data comparison,
ILES and even MDNS algorithms
can model $\alpha_b$  correctly \cite{Mue08,MueSch1_09}.

We also
present a new comparison of the differences in the spectral scaling exponents
among the three algorithm. Experiments do not provide a clear record of
RT spectral scaling exponents, but from turbulence studies \cite{Fri95}, we
expect intermittency corrections, and a steeper than -5/3 decay.
The velocity spectral properties in \cite{CabCoo06} and the
ILES simulation \cite{MorOlsWhi17} show a $-5/3$ spectral exponent.

As \cite{CabCoo06,MorOlsWhi17} employ thinly diffused initial layers
separating two fluids of distinct densities,
the immiscible experiments of \cite{SmeYou87} are the most appropriate for
comparison.
We note the very large growth of the interfacial mixing 
area, \cite{CabCoo06} Fig. 6, 
a phenomena which we have also observed \cite{LeeJinYu07,LimYuJin07}.
We believe this growth of interfacial area is a sign of a stirring
instability, as discussed next.

\begin{figure}
\begin{center}
\includegraphics[width=0.45\textwidth]{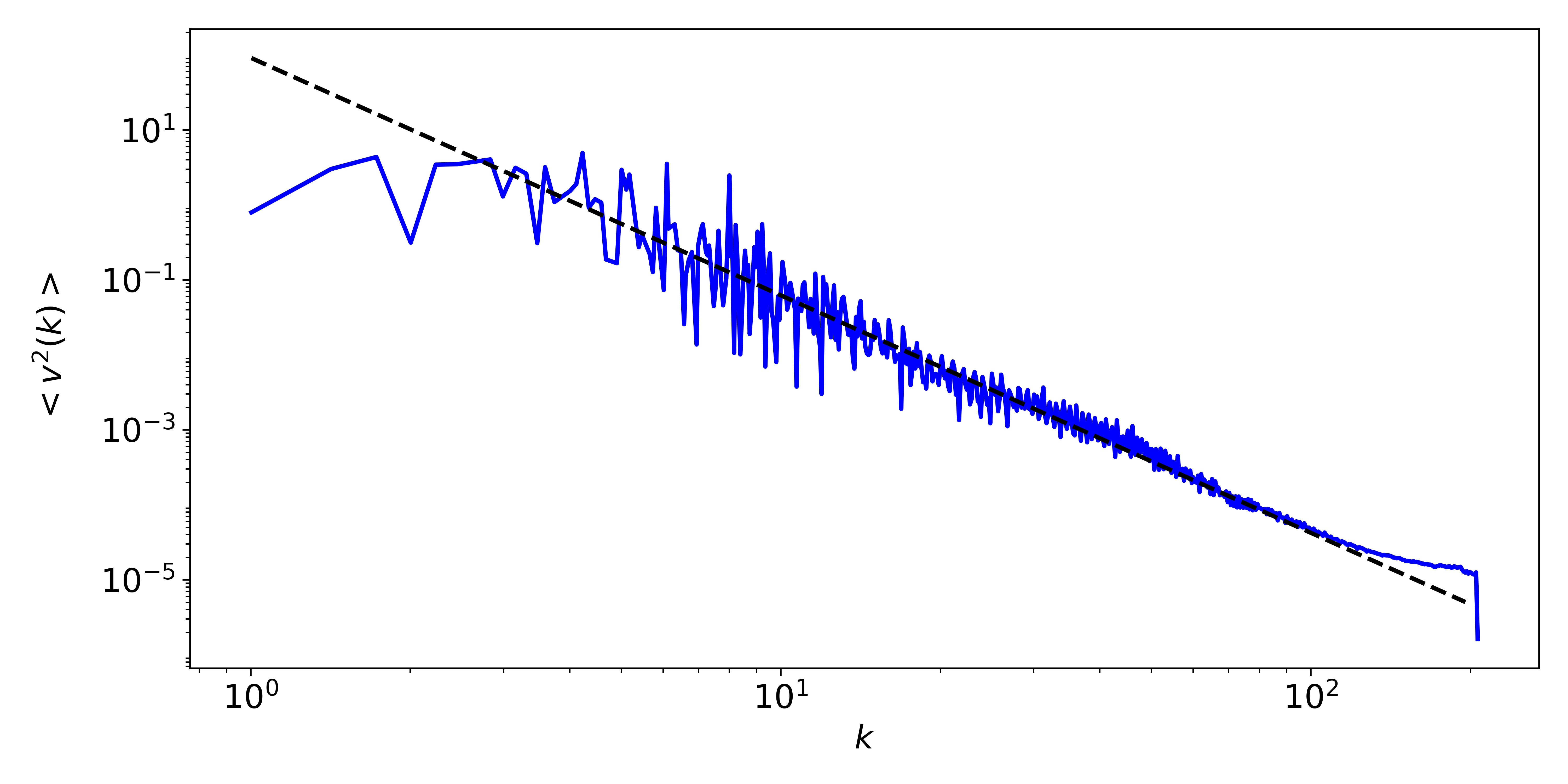}
\end{center}
\caption{
\label{fig:spectral}
Plot of the spectral decay rate, in log log variables, from the
two point function in log variables (as studied in \cite{Mah17}).
Numerical data is taken from the final time step 
RT simulations of experiment 105 reported in
\cite{ZhaKamShe18}.
The immiscible decay rate -3.17 reflects a combination of
turbulent intermittency and the effects of a stirring cascade.
}
\end{figure}

Fig.~\ref{fig:spectral} from the late time 
FronTier simulations reported in \cite{ZhaKamShe18}, shows
a strong decay rate in the velocity spectrum, 
resulting from a combination of the turbulent fractal decay 
and a separate cascading process we refer to as stirring.
Stirring is the mixing of distinct regions in a two phase flow. It occurs in the
concentration equation and is driven by velocity fluctuations. For stirring,
the concentration equation describes the
(tracked) front between the phases. Stirring fractal behavior is
less well studied than turbulent velocity. It 
accounts for the very steep velocity spectral decay seen in 
Fig.~\ref{fig:spectral}. MDNS and ILES \cite{MorOlsWhi17} capture
neither the expected turbulent intermittency correction to  the decay rate nor
any stirring correction beyond this.

\begin{table}
\caption{
\label{table:compare}
Three types of RT simulation algorithms according to the physics dissipation
principle implemented and their value for $\alpha_b$, compared
to the data of \cite{SmeYou87}.
}
\begin{centering}
\begin{tabular}{|l|l|l|l|}
\hline
Code			& energy 	& solution 	& evaluation  	\\
			& dissipation	& properties	& relative to \cite{SmeYou87} \\
\hline
\hline
MDNS			&		&			&      \\
compact high		& None		& $\alpha_b \sim 0.02$	& Inconsistent\\
order \cite{CabCoo06}	&		& 			& 	\\
\hline
Miranda			& Limited 	& $\alpha_b \sim 0.03$	& Inconsistent\\
ILES \cite{MorOlsWhi17}	&		& 			&             \\
\hline
FronTier		& Maximum	& $\alpha_b \sim 0.06$	& Consistent  \\
\cite{ZhaKamShe18}	& 		& 			&             \\
\hline
\end{tabular}
\end{centering}
\end{table}

\section{Maximum dissipation rate} 
\label{sec:p=1}

We consider a domain bounded in space 
and time for incompressible flow.
All experimental, observational and simulation data regarding turbulence
are derived from flows bounded in space and time.  We assume
such a constraint, which forces the global intensity
of the turbulence to be finite. 
According to Kolmogorov's first universality hypothesis, the fine scales,
and those
removed from spatial location near boundaries, equilibrate to steady state
isotropic fully developed turbulence.  

The idea of the proof is simple. Assuming that the global ($x,t$ integrated)
dissipation rate equals the global maximum dissipation rate,
the same must be true locally, since any difficiency of the dissipation
relative to the maximum for some
$x,t$ values on a set of positive $x,t$ measure cannot be compensated for
elsewhere.

According to Kolmogorov's theory, the rate of dissipation of energy 
in the turbulent cascade at a time $t$ is given 
by $\int_V \epsilon(x,t) \dif x$, where $V$ is the finite spatial volume
introduced above, for incompressible constant density fluids.
The integrand is nonnegative, and assuming it to be a Schwartz distribution,
the integrand is in $L_1([0,T])$.
By the same reasoning, $\epsilon(x,t) \in L_1(V)$, also a consequence of
the analysis of Sec.~\ref{sec:mdns}.
For any $T < \infty$,  the dissipated energy in the time interval $[0,T]$ is
$\int_0^T \int_V \epsilon  (x,t) \dif x \dif t$.
This is exactly the energy removed by 
viscous dissipation within the Navier-Stokes equation. We use the conventional
notation $\epsilon_0(t) = \int_V \epsilon(x,t) \dif x$.
This is the $\epsilon$
which occurs as a prefactor in (\ref{eq:K41}).
The bound on the total dissipated  energy
is our first physically motivated hypothesis.

{\bf{ Hypothesis 1.}} Finite dissipation energy and temporal reqularity.
\begin{equation}
\label{eq:linfty}
\epsilon_{(V \times [0,T])} = \int_0^T \epsilon_0(t)\dif t < \infty \ .
\end{equation}

In other words, $\epsilon_0(t) \in L_1([0,T])$.
The total energy dissipated in the interval $t \in [0,T]$, is
finite and it is limited by the total energy present at time $t = 0$, augmented
by energy supplied by a forcing term if any. We extend this hypothesis to
the (yet to be specified) maximum dissipation rate $MD(x,t)$, specifically to
$\int_V MD(x,t) \dif x = MD_0(t)$, which also has a finite integral over
$t \in [0,T]$.

{\bf{Hypothesis 2.}} Spatial regularity. For a.e. $t$, 
$\epsilon(x,t) \in L_1(V)$. That is,
\begin{equation}
\label{eq:space-linfty}
\epsilon_0(t) = \int_V \epsilon (x,t)\dif x < \infty \ .
\end{equation}
We also extend this hypothesis to $MD(x,t)$.

We add a further hypothesis, the identity of the globally
defined $\epsilon$ and $MD$. 

{\bf{ Hypothesis 3.}} Globally in space, the total energy dissipated is
the maximum possible dissipation. Thus
\begin{equation}
\label{eq:global-id}
\int_0^T \epsilon_0(t) \dif t = \int_0^T MD_0(t) \dif t \ .
\end{equation}
Dissipation, in the context of turbulent flow, refers to the passage of
energy through the turbulent cascade to progressively smaller length
scales. Hypothesis 3 states that the maximum amount of energy to be dissipated
globally in space and time is the amount which is actually dissipated globally.
This hypothesis, for non-stationary flows,
is a quasi-static interpretation of the language of K41
\cite{Kol41}. The K41 scaling laws model stationary turbulence. As observed
in \cite{Fri95}, time dependent flows may not follow the same 
(quasi-static) K41
scaling relations. Quasi-static K41 scaling laws depend for their validity on
dynamics that is sufficiently slow and smooth in time.
Specifically, we assume that the spatially averaged statistics obeys 
quasistatic K41 scaling if the high temporal frequency limit
is sufficiently weak.  RT turbulent mixing is
dominantly ${\cal{O}}(t^2)$ and appears to allow quasi-static K41 scaling.
Such an assumption is central to the definition of ILES, for example.

{\bf{Definition.}} The maximum dissipation rate of energy $MD(x,t)$ is defined
by Hypotheses 1, 2, 3  and the inequality 
\begin{equation}
\label{eq:MD}
0 \le \epsilon(x,t) \le MD(x,t) \ .
\end{equation}

{\bf Theorem.} Assume incompressible constant density flow (turbulent or not,
fully developed or not) in a periodic domain $V$. Assume 
(\ref{eq:linfty},\ref{eq:space-linfty}, \ref{eq:global-id}, \ref{eq:MD})).
Then
\begin{equation}
\label{eq:concl}
\epsilon(x,t) = MD(x,t) \in L_1(V \times [0,T]) \ ,
\end{equation}
with $T$ finite.

{\bf Proof:}
Consider the inequality
\begin{equation}
\label{eq:space-time}
0 \le  \int_0^T \int_V (MD(x,t) - \epsilon(x,t)) \dif x \dif t =  0 \ .
\end{equation}
The integrand is nonnegative for a.e. $t$ and $x$ by (\ref{eq:MD}). 
a fact which also justifies the first inequality.
The final equality is a consequence of (\ref{eq:global-id}).
It follows that the $t$ integrand, which is
the inner (space) integral, vanishes for a.e. $t$.
This means that the spatially global dissipation rates ($\epsilon$ and $MD$)
coincide, giving the RHS equality in the equation
\begin{equation}
\label{eq:space-loc}
0 \le   \int_V  (MD(x,t) - \epsilon(x,t)) \dif x  =  0 \ ,
\end{equation}
for a.e. $t$.  The lower inequality is the non-negative
integrand property, justified as before from (\ref{eq:MD}).
Thus the integrand must
vanish for a.e. $x, t$ and the proof is complete.

We next extend this proof to the case of accelerated flows, i.e.,
RT mixing. We consider two-fluid incompressible flow. The upper boundary
is now reflecting, rather than periodic. We restrict the times so that
each phase has not yet reached its opposite (top/bottom) wall. In this
case the boundary conditions, defined by reflection symmetry, are not
affected by density dependence in the gravitational or inertial forces.
In reasoning regarding total energy as a finite upper bound,
we include potential as well as kinetic energy, and make the same
hypotheses. The proof is unchanged.

\section{Significance: an example}
\label{sec:Ia}

For simulation modeling of turbulent flow nonlinearly coupled to other
physics (combustion and reactive flows, particles embedded in turbulent flow,
radiation), the method of dynamic SGS turbulent flow models, which only
deals with average subgrid effects, may be insufficient. In such cases,
the turbulent fluctuations or the full two point correlation function
is a helpful component of SGS modeling. Such a goal is only partially
realized in the simplest of cases, single density incompressible 
turbulence. For highly complex physical processes, the domain knowledge
must still be retained, and it appears to be
more feasible to bring
multifractal modeling ideas into the domain science communities.

In this spirit, we propose here a simple method
for the identification of (turbulence related) extreme events
through a modification of adaptive mesh refinement (AMR), which we 
call Fractal Mesh Refinement (FMR). We propose FMR to seek a
deflagration to detonation transition (DDT) 
in type Ia supernova. 

AMR refines the mesh where ever the algorithm detects under resolution.
In contrast, FMS skips over most of these under resolved refinements,
and only refines in those extreme cases of under resolution which are
potential candidates for DDT. By being more selective in its refinement,
FMR allows high levels of strongly focused resolution.
The method is proposed to assess the extreme events generated
by multifractal turbulent nuclear deflagration. Such events, in a white
dwarf type Ia supernova progenitor, are assumed to lead to DDT,
which produces the observed type Ia supernova.
See \cite{ZinAlmBar17,CalKruJac12} and references cited there.

FMR refines the mesh not adaptively where needed,
but only in the most highly critical regions where most important, and
thereby may detect
DDT trigger events within large volumes at a feasible computational cost.

The detailed mechanism for DDT is presumed to be diffused
radiative energy
arising from some local combustion event of extreme
intensity, in the form of a convoluted flame front, embedded in
a nearby volume of unburnt stellar material close to
ignition.
Consistent with the Zeldovich theory \cite{Lee08}, a wide spread
ignition and explosion may result. 
FMR refinement criteria will search for such
events.  See \cite{Gli18}.

There is a minimum length scale for wrinkling of a turbulent
combustion front, called the Gibson scale. Mixing can proceed in the
absence of turbulence via stirring. Thus the Gibson scale is not the
correct limiting scale for a DDT event. Stirring, for a flame front,
terminates at a smaller scale, the width of the flame itself. The analysis of
length scales must also include correctly modeled transport for charged
ions \cite{MelLimRan14},
which can be orders of magnitude larger than those inferred
from hydro considerations. The 
microstructure of mixing for a flame front could be thin flame regions
surrounded by larger regions of burned and unburned stellar material
(as with a foam of soap bubbles, with a soap film between the bubbles).
Here again multifractal and entropy issues appear to be relevant.
A multifractal clustering of smaller bubbles separated by flame fronts
can be anticipated, and
where a sufficient fraction of these bubbles are unburnt stellar material,
a trigger for DDT could occur.

This microstructure is a further law of physics, and for flame fronts, the
change of topology of the flame front occurs more frequently that would occur
in a pure stirring scenario.

For this purpose, the astrophysics code should be based on dynamic subgrid
SGS, not on ILES.

\section{Conclusions }
\label{sec:conc}

We have identified a crisis for the V\&V of numerical simulations
of turbulent or stirring flows as defined by the Euler equation.
The crisis is identified as a disputed principle of physics, in the form
of an admissibility condition, to specify the physically admissible
solution(s) of the Euler equation among the many competing solutions
which lack physical relevance. Until there is agreement over the
admissibility condition, the crisis will remain.

We have shown that the DNS claims of \cite{CabCoo06}, based on
macro flow quantities such as a globally defined Reynolds number and
Kolmogorov scale, are quite far from true DNS. We have observed
common usage of this practice, normally accompanied by a separate
validation step, missing from \cite{CabCoo06}, We introduce
a method to estimate the missing levels of mesh refinement.
We have shown that the resulting MDNS solutions are not relevant physically.

We have shown that both the MDNS and the ILES algorithms
for the solution of Euler equation 
turbulence are inadmissible physically. They are in violation of the
physical principle of maximum rate of energy dissipation, established 
mathematically here based on physically motivated hypotheses.
We have presented experimental simulation validation and physics reasoning
in support of these admissibility principles.

We have explained observations of experimental flows for which this 
error in ILES has only a minor effect. They are associated with high levels
of noise in the initial conditions, low levels of turbulent intensity,
diffusive flow parameters, and a limited choice of observables for
comparison to data.
Prior work, e.g., \cite{GliShaKam11,GeoGliLi05,ZhaKamShe18}
pertains to simulation validation studies of the RT instability
experiments \cite{SmeYou87} with a stronger intensity of turbulence
and for which such significant long wave length perturbations to the
initial data are missing. In these experiments, the present analysis 
provides a partial explanation for the factor of about 2 discrepancy between
observed and ILES predicted instability growth rates. Long wave length
noise in the initial conditions
has been ruled out as an explanation for this discrepancy.

We have noted the potential for ILES related errors to influence
ongoing scientific investigations, including the search for
DDT in type Ia supernova.

Clearly V\&V standards should include an analysis of the physical
relevance of proposed solutions to flow problems, specifically turbulent
and stirring problems.
The ILES simulations of the experiments of \cite{SmeYou87} fail this
test by a factor of 2 in the RT growth rate $\alpha_b$, and
on this basis we judge them to be physically inadmissible. Likewise
the MDNS (claimed DNS) solutions of \cite{CabCoo06}, which differ from
experiment by a larger factor, are judged to be inadmissible.

We recognize that the conclusions of this paper will be controversial within 
the ILES and high order compact turbulent simulation communities. 
A deeper consideration of the
issues raised here is a possible outcome. 
The issues to be analyzed are clear:
\begin{itemize}
\item
Is the transport of energy and concentration, blocked at the
grid level, to be ignored entirely \cite{CabCoo06}? 
Are the full standards of DNS simulations to be ignored in simulations
claiming to be DNS? If MDNS is used, is there a need for a separate 
validation step? 
\item
Is the energy dissipation
to be regarded as a Gibbs phenomena \cite{MorOlsWhi17},
and thus to be minimized? 
\item
Is energy dissipation a physical phenomena,
to be modeled accurately \cite{GerPioMoi91,MoiSquCab91}?
\end{itemize}

If the response to this paper is an appeal to 
consensus (everyone else is doing it),
the argument fails. Consensus is of course a weak argument, and
one that flies  in the face of standards of V\&V. More significantly,
there is a far larger engineering community
using dynamic SGS models in the design of engineering structures
tested in actual practice.
This choice is backed by nearly three decades of
extensive experimental validation.  It is further used
to extend the calibration range of RANS simulations beyond available
experimental data. The resulting RANS, calibrated to dynamic SGS LES data,
are even more widely used in the design and optimization
of engineering structures; these are also tested in real applications.
Consensus in this larger community overwhelms the ILES consensus
by its shear magnitude and by its nearly three decades
of designed structures that
are ``live tested'' in actual operations. ILES loses the consensus argument.

\section*{Acknowledgements}
\label{ack}

Use of computational support by the Swiss National Supercomputing Centre is gratefully acknowledged. 
Los Alamos National Laboratory Preprint LA-UR-19-20285.


\begin{thebibliography}{38}
\expandafter\ifx\csname natexlab\endcsname\relax\def\natexlab#1{#1}\fi
\expandafter\ifx\csname bibnamefont\endcsname\relax
  \def\bibnamefont#1{#1}\fi
\expandafter\ifx\csname bibfnamefont\endcsname\relax
  \def\bibfnamefont#1{#1}\fi
\expandafter\ifx\csname citenamefont\endcsname\relax
  \def\citenamefont#1{#1}\fi
\expandafter\ifx\csname url\endcsname\relax
  \def\url#1{\texttt{#1}}\fi
\expandafter\ifx\csname urlprefix\endcsname\relax\def\urlprefix{URL }\fi
\providecommand{\bibinfo}[2]{#2}
\providecommand{\eprint}[2][]{\url{#2}}

\bibitem[{\citenamefont{{De Leliss} and Szekelyhidi}(2009)}]{DelSze09}
\bibinfo{author}{\bibfnamefont{C.}~\bibnamefont{{De Leliss}}} \bibnamefont{and}
  \bibinfo{author}{\bibfnamefont{L.}~\bibnamefont{Szekelyhidi}},
  \bibinfo{journal}{Ann. Math.} \textbf{\bibinfo{volume}{170}},
  \bibinfo{pages}{1471} (\bibinfo{year}{2009}).

\bibitem[{\citenamefont{{De Leliss} and Szekelyhidi}(2010)}]{DelSze10}
\bibinfo{author}{\bibfnamefont{C.}~\bibnamefont{{De Leliss}}} \bibnamefont{and}
  \bibinfo{author}{\bibfnamefont{L.}~\bibnamefont{Szekelyhidi}},
  \bibinfo{journal}{Arch. Rat. Mech. Anal.} \textbf{\bibinfo{volume}{195}},
  \bibinfo{pages}{225} (\bibinfo{year}{2010}).

\bibitem[{\citenamefont{Zhou}(2017{\natexlab{a}})}]{Zho17a}
\bibinfo{author}{\bibfnamefont{Y.}~\bibnamefont{Zhou}},
  \bibinfo{journal}{Physics Reports} \textbf{\bibinfo{volume}{720--722}},
  \bibinfo{pages}{1} (\bibinfo{year}{2017}{\natexlab{a}}),
  \bibinfo{note}{http://dx.doi.org/10.1016/j.physrep.2017.07.005}.

\bibitem[{\citenamefont{Zhou}(2017{\natexlab{b}})}]{Zho17b}
\bibinfo{author}{\bibfnamefont{Y.}~\bibnamefont{Zhou}},
  \bibinfo{journal}{Physics Reports} \textbf{\bibinfo{volume}{723--725}},
  \bibinfo{pages}{1} (\bibinfo{year}{2017}{\natexlab{b}}),
  \bibinfo{note}{http://doi.org/10.1016/j.physrep.2017.07.008}.

\bibitem[{\citenamefont{Smeeton and Youngs}(1987)}]{SmeYou87}
\bibinfo{author}{\bibfnamefont{V.~S.} \bibnamefont{Smeeton}} \bibnamefont{and}
  \bibinfo{author}{\bibfnamefont{D.~L.} \bibnamefont{Youngs}},
  \bibinfo{type}{AWE Report Number} \bibinfo{number}{0 35/87}
  (\bibinfo{year}{1987}).

\bibitem[{\citenamefont{Mueschke and Schilling}(2009)}]{MueSch1_09}
\bibinfo{author}{\bibfnamefont{N.}~\bibnamefont{Mueschke}} \bibnamefont{and}
  \bibinfo{author}{\bibfnamefont{O.}~\bibnamefont{Schilling}},
  \bibinfo{journal}{Physics of Fluids} \textbf{\bibinfo{volume}{21}},
  \bibinfo{pages}{014106 1} (\bibinfo{year}{2009}).

\bibitem[{\citenamefont{Lim et~al.}(2012)\citenamefont{Lim, Kaman, Yu, Mahadeo,
  Xu, Zhang, Glimm, Dutta, Sharp, and Plohr}}]{LimKamYu12}
\bibinfo{author}{\bibfnamefont{H.}~\bibnamefont{Lim}},
  \bibinfo{author}{\bibfnamefont{T.}~\bibnamefont{Kaman}},
  \bibinfo{author}{\bibfnamefont{Y.}~\bibnamefont{Yu}},
  \bibinfo{author}{\bibfnamefont{V.}~\bibnamefont{Mahadeo}},
  \bibinfo{author}{\bibfnamefont{Y.}~\bibnamefont{Xu}},
  \bibinfo{author}{\bibfnamefont{H.}~\bibnamefont{Zhang}},
  \bibinfo{author}{\bibfnamefont{J.}~\bibnamefont{Glimm}},
  \bibinfo{author}{\bibfnamefont{S.}~\bibnamefont{Dutta}},
  \bibinfo{author}{\bibfnamefont{D.~H.} \bibnamefont{Sharp}}, \bibnamefont{and}
  \bibinfo{author}{\bibfnamefont{B.}~\bibnamefont{Plohr}},
  \bibinfo{journal}{Acta Mathematica Scientia} \textbf{\bibinfo{volume}{32}},
  \bibinfo{pages}{237} (\bibinfo{year}{2012}), \bibinfo{note}{{S}tony Brook
  University Preprint SUNYSB-AMS-11-07 and Los Alamos National Laboratory
  Preprint LA-UR 11-05862}.

\bibitem[{\citenamefont{George et~al.}(2006)\citenamefont{George, Glimm, Li,
  Li, and Liu}}]{GeoGliLi05}
\bibinfo{author}{\bibfnamefont{E.}~\bibnamefont{George}},
  \bibinfo{author}{\bibfnamefont{J.}~\bibnamefont{Glimm}},
  \bibinfo{author}{\bibfnamefont{X.-L.} \bibnamefont{Li}},
  \bibinfo{author}{\bibfnamefont{Y.-H.} \bibnamefont{Li}}, \bibnamefont{and}
  \bibinfo{author}{\bibfnamefont{X.-F.} \bibnamefont{Liu}},
  \bibinfo{journal}{Phys. Rev. E} \textbf{\bibinfo{volume}{73}},
  \bibinfo{pages}{016304} (\bibinfo{year}{2006}).

\bibitem[{\citenamefont{Glimm et~al.}(2013)\citenamefont{Glimm, Sharp, Kaman,
  and Lim}}]{GliShaKam11}
\bibinfo{author}{\bibfnamefont{J.}~\bibnamefont{Glimm}},
  \bibinfo{author}{\bibfnamefont{D.~H.} \bibnamefont{Sharp}},
  \bibinfo{author}{\bibfnamefont{T.}~\bibnamefont{Kaman}}, \bibnamefont{and}
  \bibinfo{author}{\bibfnamefont{H.}~\bibnamefont{Lim}},
  \bibinfo{journal}{Phil. Trans. R. Soc. A} \textbf{\bibinfo{volume}{371}},
  \bibinfo{pages}{20120183} (\bibinfo{year}{2013}), \bibinfo{note}{{L}os Alamos
  National Laboratory Preprint LA-UR 11-00423 and Stony Brook University
  Preprint SUNYSB-AMS-11-01}.

\bibitem[{\citenamefont{Zhang et~al.}(2018)\citenamefont{Zhang, Kaman, She,
  Cheng, Glimm, and Sharp}}]{ZhaKamShe18}
\bibinfo{author}{\bibfnamefont{H.}~\bibnamefont{Zhang}},
  \bibinfo{author}{\bibfnamefont{T.}~\bibnamefont{Kaman}},
  \bibinfo{author}{\bibfnamefont{D.}~\bibnamefont{She}},
  \bibinfo{author}{\bibfnamefont{B.}~\bibnamefont{Cheng}},
  \bibinfo{author}{\bibfnamefont{J.}~\bibnamefont{Glimm}}, \bibnamefont{and}
  \bibinfo{author}{\bibfnamefont{D.~H.} \bibnamefont{Sharp}},
  \bibinfo{journal}{Pure and Applied Mathematics Quarterly}
  (\bibinfo{year}{2018}), \bibinfo{note}{in press; Los Alamos National
  Laboratory preprint LA-UR-18-22134}.

\bibitem[{\citenamefont{Lanford}(1973)}]{Lan73}
\bibinfo{author}{\bibfnamefont{O.}~\bibnamefont{Lanford}},
  \bibinfo{journal}{Lecture Notes in Physics 20}  (\bibinfo{year}{1973}).

\bibitem[{\citenamefont{Lebowitz}(1978)}]{Leb78}
\bibinfo{author}{\bibfnamefont{J.~L.} \bibnamefont{Lebowitz}},
  \bibinfo{journal}{Prog. Theor. Phys. suppl.} \textbf{\bibinfo{volume}{20}}
  (\bibinfo{year}{1978}).

\bibitem[{\citenamefont{Martyushev and Seleznev}(2006)}]{MarSel06}
\bibinfo{author}{\bibfnamefont{L.~M.} \bibnamefont{Martyushev}}
  \bibnamefont{and} \bibinfo{author}{\bibfnamefont{V.~D.}
  \bibnamefont{Seleznev}}, \bibinfo{journal}{Phy. Reports}
  \textbf{\bibinfo{volume}{426}}, \bibinfo{pages}{1} (\bibinfo{year}{2006}).

\bibitem[{\citenamefont{Mihelich et~al.}(2017)\citenamefont{Mihelich, Faranda,
  Pailard, and Dubrulle}}]{MihFarPai17}
\bibinfo{author}{\bibfnamefont{M.}~\bibnamefont{Mihelich}},
  \bibinfo{author}{\bibfnamefont{D.}~\bibnamefont{Faranda}},
  \bibinfo{author}{\bibfnamefont{D.}~\bibnamefont{Pailard}}, \bibnamefont{and}
  \bibinfo{author}{\bibfnamefont{B.}~\bibnamefont{Dubrulle}},
  \bibinfo{journal}{Entropy} \textbf{\bibinfo{volume}{19}}
  (\bibinfo{year}{2017}).

\bibitem[{\citenamefont{Ozawa et~al.}(2003)\citenamefont{Ozawa, Ohmura,
  Lorentz, and Pujol}}]{OzaOhmLor03}
\bibinfo{author}{\bibfnamefont{H.}~\bibnamefont{Ozawa}},
  \bibinfo{author}{\bibfnamefont{A.}~\bibnamefont{Ohmura}},
  \bibinfo{author}{\bibfnamefont{R.}~\bibnamefont{Lorentz}}, \bibnamefont{and}
  \bibinfo{author}{\bibfnamefont{T.}~\bibnamefont{Pujol}},
  \bibinfo{journal}{Reviews of Geophysics} \textbf{\bibinfo{volume}{41}}
  (\bibinfo{year}{2003}).

\bibitem[{Kle(2010)}]{KleDyk10}
in \emph{\bibinfo{booktitle}{What is Maximum Entrpy Productionn and how should
  we apply it}}, edited by
  \bibinfo{editor}{\bibfnamefont{A.}~\bibnamefont{Kleidon}} \bibnamefont{and}
  \bibinfo{editor}{\bibfnamefont{J.}~\bibnamefont{Dyke}}
  (\bibinfo{publisher}{Entropy}, \bibinfo{year}{2010}), \bibinfo{note}{special
  issue, Vol. 12}.

\bibitem[{\citenamefont{Ge and Qian}(2009)}]{GeQia09}
\bibinfo{author}{\bibfnamefont{H.}~\bibnamefont{Ge}} \bibnamefont{and}
  \bibinfo{author}{\bibfnamefont{H.}~\bibnamefont{Qian}},
  \bibinfo{journal}{ArXive:0911.3984v2 cond-mat.stat-mech}
  (\bibinfo{year}{2009}).

\bibitem[{\citenamefont{Germano et~al.}(1991)\citenamefont{Germano, Piomelli,
  Moin, and Cabot}}]{GerPioMoi91}
\bibinfo{author}{\bibfnamefont{M.}~\bibnamefont{Germano}},
  \bibinfo{author}{\bibfnamefont{U.}~\bibnamefont{Piomelli}},
  \bibinfo{author}{\bibfnamefont{P.}~\bibnamefont{Moin}}, \bibnamefont{and}
  \bibinfo{author}{\bibfnamefont{W.~H.} \bibnamefont{Cabot}},
  \bibinfo{journal}{Phys. Fluids A} \textbf{\bibinfo{volume}{3}},
  \bibinfo{pages}{1760} (\bibinfo{year}{1991}).

\bibitem[{\citenamefont{Moin et~al.}(1991)\citenamefont{Moin, Squires, Cabot,
  and Lee}}]{MoiSquCab91}
\bibinfo{author}{\bibfnamefont{P.}~\bibnamefont{Moin}},
  \bibinfo{author}{\bibfnamefont{K.}~\bibnamefont{Squires}},
  \bibinfo{author}{\bibfnamefont{W.}~\bibnamefont{Cabot}}, \bibnamefont{and}
  \bibinfo{author}{\bibfnamefont{S.}~\bibnamefont{Lee}},
  \bibinfo{journal}{Phys. Fluids A} \textbf{\bibinfo{volume}{3}},
  \bibinfo{pages}{2746} (\bibinfo{year}{1991}).

\bibitem[{\citenamefont{Kolmogorov}(1941)}]{Kol41}
\bibinfo{author}{\bibfnamefont{A.~N.} \bibnamefont{Kolmogorov}},
  \bibinfo{journal}{Doklady Akad. Nauk. SSSR} \textbf{\bibinfo{volume}{30}},
  \bibinfo{pages}{299} (\bibinfo{year}{1941}).

\bibitem[{\citenamefont{Morgan et~al.}(2017)\citenamefont{Morgan, Olson, White,
  and McFarland}}]{MorOlsWhi17}
\bibinfo{author}{\bibfnamefont{B.~E.} \bibnamefont{Morgan}},
  \bibinfo{author}{\bibfnamefont{B.~J.} \bibnamefont{Olson}},
  \bibinfo{author}{\bibfnamefont{J.~E.} \bibnamefont{White}}, \bibnamefont{and}
  \bibinfo{author}{\bibfnamefont{J.~A.} \bibnamefont{McFarland}},
  \bibinfo{journal}{J. Turbulence} \textbf{\bibinfo{volume}{18}}
  (\bibinfo{year}{2017}).

\bibitem[{\citenamefont{Cabot and Cook}(2006)}]{CabCoo06}
\bibinfo{author}{\bibfnamefont{W.}~\bibnamefont{Cabot}} \bibnamefont{and}
  \bibinfo{author}{\bibfnamefont{A.}~\bibnamefont{Cook}},
  \bibinfo{journal}{Nature Physics} \textbf{\bibinfo{volume}{2}},
  \bibinfo{pages}{562} (\bibinfo{year}{2006}).

\bibitem[{\citenamefont{Kandea and Morishita}(2007)}]{KanMor07}
\bibinfo{author}{\bibfnamefont{Y.}~\bibnamefont{Kandea}} \bibnamefont{and}
  \bibinfo{author}{\bibfnamefont{K.}~\bibnamefont{Morishita}},
  \bibinfo{journal}{J. Phys. Soc. Japan} \textbf{\bibinfo{volume}{7}}
  (\bibinfo{year}{2007}).

\bibitem[{\citenamefont{Sawford and Yeung}(2015)}]{SawYeu15}
\bibinfo{author}{\bibfnamefont{B.}~\bibnamefont{Sawford}} \bibnamefont{and}
  \bibinfo{author}{\bibfnamefont{P.~K.} \bibnamefont{Yeung}},
  \textbf{\bibinfo{volume}{27}} (\bibinfo{year}{2015}).

\bibitem[{\citenamefont{Moin and Mahesh}(1998)}]{MoiMah98}
\bibinfo{author}{\bibfnamefont{P.}~\bibnamefont{Moin}} \bibnamefont{and}
  \bibinfo{author}{\bibfnamefont{K.}~\bibnamefont{Mahesh}},
  \bibinfo{journal}{Ann. Rev. Fluid Mech.} \textbf{\bibinfo{volume}{30}},
  \bibinfo{pages}{539} (\bibinfo{year}{1998}).

\bibitem[{\citenamefont{Rehagen et~al.}(2017)\citenamefont{Rehagen, Greenough,
  and Olson}}]{RehGreOls17}
\bibinfo{author}{\bibfnamefont{T.}~\bibnamefont{Rehagen}},
  \bibinfo{author}{\bibfnamefont{J.}~\bibnamefont{Greenough}},
  \bibnamefont{and} \bibinfo{author}{\bibfnamefont{B.}~\bibnamefont{Olson}},
  \bibinfo{journal}{J. Fluids Eng.} \textbf{\bibinfo{volume}{139}}
  (\bibinfo{year}{2017}).

\bibitem[{\citenamefont{Kolmogorov}(1962)}]{Kol62}
\bibinfo{author}{\bibfnamefont{A.~N.} \bibnamefont{Kolmogorov}},
  \bibinfo{journal}{J. Fluid Mechanics} \textbf{\bibinfo{volume}{13}},
  \bibinfo{pages}{82} (\bibinfo{year}{1962}).

\bibitem[{\citenamefont{Lim et~al.}(2010)\citenamefont{Lim, Iwerks, Glimm, and
  Sharp}}]{LimIweGli10}
\bibinfo{author}{\bibfnamefont{H.}~\bibnamefont{Lim}},
  \bibinfo{author}{\bibfnamefont{J.}~\bibnamefont{Iwerks}},
  \bibinfo{author}{\bibfnamefont{J.}~\bibnamefont{Glimm}}, \bibnamefont{and}
  \bibinfo{author}{\bibfnamefont{D.~H.} \bibnamefont{Sharp}},
  \bibinfo{journal}{Proc. Natl. Acad. Sci.} \textbf{\bibinfo{volume}{107(29)}},
  \bibinfo{pages}{12786} (\bibinfo{year}{2010}), \bibinfo{note}{{S}tony Brook
  University Preprint SUNYSB-AMS-09-05 and Los Alamos National Laboratory
  Preprint LA-UR 09-06333}.

\bibitem[{\citenamefont{Mueschke}(2008)}]{Mue08}
\bibinfo{author}{\bibfnamefont{N.~J.} \bibnamefont{Mueschke}}, Ph.D. thesis,
  \bibinfo{school}{Texas A and M University} (\bibinfo{year}{2008}).

\bibitem[{\citenamefont{Frisch}(1996)}]{Fri95}
\bibinfo{author}{\bibfnamefont{U.}~\bibnamefont{Frisch}},
  \emph{\bibinfo{title}{Turbulence: The Legacy of {A}. {N}. {K}olmogorov}}
  (\bibinfo{publisher}{Cambridge Univeristy Press},
  \bibinfo{address}{Cambridge}, \bibinfo{year}{1996}).

\bibitem[{\citenamefont{Lee et~al.}(2008)\citenamefont{Lee, Jin, Yu, and
  Glimm}}]{LeeJinYu07}
\bibinfo{author}{\bibfnamefont{H.}~\bibnamefont{Lee}},
  \bibinfo{author}{\bibfnamefont{H.}~\bibnamefont{Jin}},
  \bibinfo{author}{\bibfnamefont{Y.}~\bibnamefont{Yu}}, \bibnamefont{and}
  \bibinfo{author}{\bibfnamefont{J.}~\bibnamefont{Glimm}},
  \bibinfo{journal}{Phys. Fluids} \textbf{\bibinfo{volume}{20}},
  \bibinfo{pages}{1} (\bibinfo{year}{2008}), \bibinfo{note}{{S}tony {B}rook
  {U}niversity Preprint SUNYSB-AMS-07-03}.

\bibitem[{\citenamefont{Lim et~al.}(2008)\citenamefont{Lim, Yu, Jin, Kim, Lee,
  Glimm, Li, and Sharp}}]{LimYuJin07}
\bibinfo{author}{\bibfnamefont{H.}~\bibnamefont{Lim}},
  \bibinfo{author}{\bibfnamefont{Y.}~\bibnamefont{Yu}},
  \bibinfo{author}{\bibfnamefont{H.}~\bibnamefont{Jin}},
  \bibinfo{author}{\bibfnamefont{D.}~\bibnamefont{Kim}},
  \bibinfo{author}{\bibfnamefont{H.}~\bibnamefont{Lee}},
  \bibinfo{author}{\bibfnamefont{J.}~\bibnamefont{Glimm}},
  \bibinfo{author}{\bibfnamefont{X.-L.} \bibnamefont{Li}}, \bibnamefont{and}
  \bibinfo{author}{\bibfnamefont{D.~H.} \bibnamefont{Sharp}},
  \bibinfo{journal}{Compu. Methods Appl. Mech. Engrg.}
  \textbf{\bibinfo{volume}{197}}, \bibinfo{pages}{3435} (\bibinfo{year}{2008}),
  \bibinfo{note}{{S}tony Brook University Preprint SUNYSB-AMS-07-05}.

\bibitem[{\citenamefont{Mahadeo}(2017)}]{Mah17}
\bibinfo{author}{\bibfnamefont{V.}~\bibnamefont{Mahadeo}}, \bibinfo{type}{Ph.d.
  thesis}, \bibinfo{school}{Stony Brook University} (\bibinfo{year}{2017}).

\bibitem[{\citenamefont{Zingale et~al.}(2017)\citenamefont{Zingale, Almgren,
  Sazo, Beckner, Bell, Friesen, Jacobs, Katz, Malone, Nonaka
  et~al.}}]{ZinAlmBar17}
\bibinfo{author}{\bibfnamefont{M.}~\bibnamefont{Zingale}},
  \bibinfo{author}{\bibfnamefont{A.~S.} \bibnamefont{Almgren}},
  \bibinfo{author}{\bibfnamefont{M.~G.~B.} \bibnamefont{Sazo}},
  \bibinfo{author}{\bibfnamefont{V.~E.} \bibnamefont{Beckner}},
  \bibinfo{author}{\bibfnamefont{J.~B.} \bibnamefont{Bell}},
  \bibinfo{author}{\bibfnamefont{B.}~\bibnamefont{Friesen}},
  \bibinfo{author}{\bibfnamefont{A.~M.} \bibnamefont{Jacobs}},
  \bibinfo{author}{\bibfnamefont{M.}~\bibnamefont{Katz}},
  \bibinfo{author}{\bibfnamefont{C.~M.} \bibnamefont{Malone}},
  \bibinfo{author}{\bibfnamefont{A.~J.} \bibnamefont{Nonaka}},
  \bibnamefont{et~al.}, \bibinfo{journal}{ArXive: 1771-06203}
  (\bibinfo{year}{2017}).

\bibitem[{\citenamefont{Calder et~al.}(2012)\citenamefont{Calder, Krueger,
  Jackson, Townsley, Brown, and Times}}]{CalKruJac12}
\bibinfo{author}{\bibfnamefont{A.}~\bibnamefont{Calder}},
  \bibinfo{author}{\bibfnamefont{B.}~\bibnamefont{Krueger}},
  \bibinfo{author}{\bibfnamefont{A.}~\bibnamefont{Jackson}},
  \bibinfo{author}{\bibfnamefont{D.}~\bibnamefont{Townsley}},
  \bibinfo{author}{\bibfnamefont{E.}~\bibnamefont{Brown}}, \bibnamefont{and}
  \bibinfo{author}{\bibfnamefont{F.}~\bibnamefont{Times}},
  \bibinfo{journal}{arXive: 1205-0966}  (\bibinfo{year}{2012}).

\bibitem[{\citenamefont{Lee}(2008)}]{Lee08}
\bibinfo{author}{\bibfnamefont{J.}~\bibnamefont{Lee}},
  \emph{\bibinfo{title}{The Detonation Phenomena}}
  (\bibinfo{publisher}{Cambridge University Press}, \bibinfo{year}{2008}).

\bibitem[{\citenamefont{Glimm}(2018)}]{Gli18}
\bibinfo{author}{\bibfnamefont{J.}~\bibnamefont{Glimm}},
  \bibinfo{journal}{arXive:1205-0966}  (\bibinfo{year}{2018}),
  \bibinfo{note}{1806 06054}.

\bibitem[{\citenamefont{Melvin et~al.}(2015)\citenamefont{Melvin, Lim, Rana,
  Cheng, Glimm, Sharp, and Wilson}}]{MelLimRan14}
\bibinfo{author}{\bibfnamefont{J.}~\bibnamefont{Melvin}},
  \bibinfo{author}{\bibfnamefont{H.}~\bibnamefont{Lim}},
  \bibinfo{author}{\bibfnamefont{V.}~\bibnamefont{Rana}},
  \bibinfo{author}{\bibfnamefont{B.}~\bibnamefont{Cheng}},
  \bibinfo{author}{\bibfnamefont{J.}~\bibnamefont{Glimm}},
  \bibinfo{author}{\bibfnamefont{D.~H.} \bibnamefont{Sharp}}, \bibnamefont{and}
  \bibinfo{author}{\bibfnamefont{D.~C.} \bibnamefont{Wilson}},
  \bibinfo{journal}{Physics of Plasmas} \textbf{\bibinfo{volume}{22}},
  \bibinfo{pages}{022708} (\bibinfo{year}{2015}).

\end{thebibliography}
\end{document}